\documentclass[pre,aps,twocolumn]{revtex4}

\usepackage{graphicx}
\usepackage{amsmath}
\usepackage{amssymb}
\usepackage{ifthen}
\usepackage{booktabs}

\usepackage{graphicx}% Include figure files
\usepackage{dcolumn}% Align table columns on decimal point
\usepackage{bm}% bold math
\usepackage{longtable}
\usepackage{gensymb}
\usepackage{color}

\newcommand{\bb}{\begin{equation}}
\newcommand{\ee}{\end{equation}}
\newcommand{\ba}{\begin{eqnarray*}}
\newcommand{\ea}{\end{eqnarray*}}
\newcommand{\rhor}{\rho({\bf r})}
\newcommand{\dd}{{\rm d}}
\newcommand{\rr}{{\mathbf r}}
\newcommand{\dr}{{\rm d}{\bf r}}

\begin{document}

\title{Bridging of liquid drops at chemically structured walls}

\author{Alexandr \surname{Malijevsk\'y}}
\affiliation{
{Department of Physical Chemistry, University of Chemical Technology Prague, Praha 6, 166 28, Czech Republic;}\\
 {Department of Molecular and Mesoscopic Modelling, ICPF of the Czech Academy Sciences, Prague, Czech Republic}}                %Institute of Chemical Process Fundamentals of the CAS, v. v. i., Prague, Czech Republic}}
\author{Andrew O. \surname{Parry}}
\affiliation{Department of Mathematics, Imperial College London, London SW7 2BZ, UK}
\author{Martin \surname{Posp\'\i\v sil}}
\affiliation{{Department of Physical Chemistry, University of Chemical Technology Prague, Praha 6, 166 28, Czech Republic;}\\
 {Department of Molecular and Mesoscopic Modelling, ICPF of the Czech Academy Sciences, Prague, Czech Republic}}

\begin{abstract}
\noindent Using mesoscopic interfacial models and microscopic density functional theory we study fluid adsorption at a dry wall decorated with three
completely wet stripes of width $L$ separated by distances $D_1$ and $D_2$. The stripes interact with the fluid with long-range forces inducing a
large finite-size contribution to the surface free-energy. We show that this non-extensive free-energy contribution scales with $\ln L$ and drives
different types of bridging transition corresponding to the merging of liquid drops adsorbed at neighbouring wetting stripes when the separation
between them is \emph{molecularly} small. We determine the surface phase diagram and show that this exhibits two triple points, where isolated drops,
double drops and triple drops coexist. For the symmetric case, $D_1=D_2\equiv D$, our results also confirm that the equilbrium droplet configuration
always has the symmetry of the substrate corresponding to either three isolated drops when $D$ is large or a single triple drop when $D$ is small;
however, symmetry broken configurations do occur in a metastable part of the phase diagram which lies very close to the equilibrium bridging phase
boundary. Implications for phase transitions on other types of patterned surface are considered.
\end{abstract}

\maketitle

\section{Introduction}

The study of fluid adsorption and droplet formation on structured and patterned surfaces has received a great deal of attention in recent years. This
is in part motivated by practical applications such as microfluidics \cite{chou} and superhydrophobicity \cite{quere1, quere2} and also, more
fundamentally, by the many new types of interfacial phase transitions \cite{napior, parry2000, abraham, milchev, wedge_prl, wedge_mal, carlos,
delfino, groove_tas, bruschi, groove_parry, groove_mal, stewart, nold, singh, giacomello} that can occur compared to those at structureless and
chemically homogeneous surfaces \cite{dietrich, schick, forgacs, saam}. Consider, for example, a flat substrate (wall) decorated with one or more
stripes which have an enhanced wettability; that is a lower contact angle $\theta$, compared to the rest of the wall, which serves to preferentially
nucleate liquid. When the volume of liquid is fixed, such surfaces can induce macroscopic morphological phase transitions  associated with the local
breaking of translational invariance and Young's law \cite{lenz, gau}. In the grand canonical ensemble, that is when the volume of liquid is not
constrained, morphological transitions do not occur since the fluid density profile must have the same symmetry as the confining external field
induced by the wall. However in their place are a number of phase transitions including the possibility of the formation of liquid bridges that span
between different striped regions when the gaps between them are sufficiently small \cite{mpp}. These are similar to bridging transitions, that is
the local condensation of liquid, between nano-particles (spheres, cylinders etc.)  immersed in a solvent reservoir \cite{dobbs, bauer, hopkins,
mal15, malpar15}.

In the present paper we study bridging transitions on a chemically heterogeneous surface decorated with three identical stripes of width $L$ that are
completely wet (contact angle $\theta=0$). This topography allows, in principle, for stable and metastable droplet and bridging configurations which
either follow or break the symmetry of the substrate. Our motivation is not only to show that bridging transitions must occur as the distance between
the stripes is reduced but also that these occur in such a way as to suppress the possibility of equilibrium spontaneous symmetry breaking associated
with different bridge-like coverings of the wall which only occur in a metastable part of the phase diagram. However, as we will see the quantitative
nature of the metastability  is quite surprising with repercussions for the structure of the phase diagram. Our paper is arranged as follows; in
section II we use simple scaling and interfacial Hamiltonian theory to predict equilibrium droplet configurations on a dry surface decorated with
one, two and three stripes which are completely wet. We focus on systems with long ranged dispersion-like forces for which there are important
finite-size contributions to the surface free-energy of drop configurations. For a surface with three stripes we distinguish between symmetric and
broken symmetric droplet shapes and determine the location of different possible phase transitions between them. In this way, we are able to predict
the phase diagram when we vary the distances $D_1$ and $D_2$ between the stripes. In section III we confirm these predictions using microscopic
density functional theory where we specialise to a perfectly dry wall patterned with wet stripes. We finish with a discussion concerning
generalizations and repercussions for phase transitions on periodically decorated substrates.

\section{Interface Hamiltonian model}

To begin, we recall briefly the features of a drop nucleated at a single stripe and in particular why, for systems with dispersion-like forces, there
is a logarithmic contribution to the excess surface free-energy. This is crucial when considering the quantitative aspects of the different bridging
phase boundaries when the substrate is decorated with more stripes. Consider a planar wall in contact with vapour at chemical potential $\mu$ close
to saturation $\mu_{\rm sat}$ at temperature $T$ far below the bulk critical value $T_c$. The wall is partially wet by liquid, corresponding to
contact angle $\theta$, except along a macroscopically long, deep stripe of width $L$ which is made of a different material which is completely wet.
The stripe nucleates a liquid drop which remains of finite height $h_m$, even at saturation chemical potential, which depends on the stripe width.
Simple finite-size scaling arguments predict that at saturation $h_m\propto L^{\beta_s^{\rm co}/\nu_\parallel^{\rm co}}$ where $\beta_s^{\rm co}$ and
$\nu_\parallel^{\rm co}$ are, respectively, the critical exponents for the adsorption and parallel correlation length characterizing a complete
wetting phase transition \cite{dietrich, schick, forgacs, mpp, conf}. Similarly the surface free-energy per unit length of the stripe should contain
a singular contribution $f_{s}\propto L^{1-(2-\alpha_s^{\rm co})/\nu_\parallel^{\rm co}}$ where now $\alpha_s^{\rm co}$ is the surface specific heat
exponent. For systems with long-ranged dispersion forces the upper critical dimension for complete wetting $d^*_{co}<3$, so that in three dimensions
the mean-field values of the critical exponents $\alpha_s^{\rm co}=4/3$, $\beta_s^{\rm co}=1/3$ and $\nu_\parallel^{\rm co}=2/3$ are exact
\cite{lip_ucd}. This predicts that the droplet height scales as $h_{m}\propto \sqrt{L}$, while the exponent for the free-energy {\it{vanishes}}
suggesting that this is a marginal case allowing for logarithmic corrections. These scaling predictions contrast with the situation for systems with
short-ranged forces for which we anticipate $h_m\propto\ln L$ and $f_s\propto 1/L^2$.

\begin{figure*}
\centerline{\includegraphics[width=12cm]{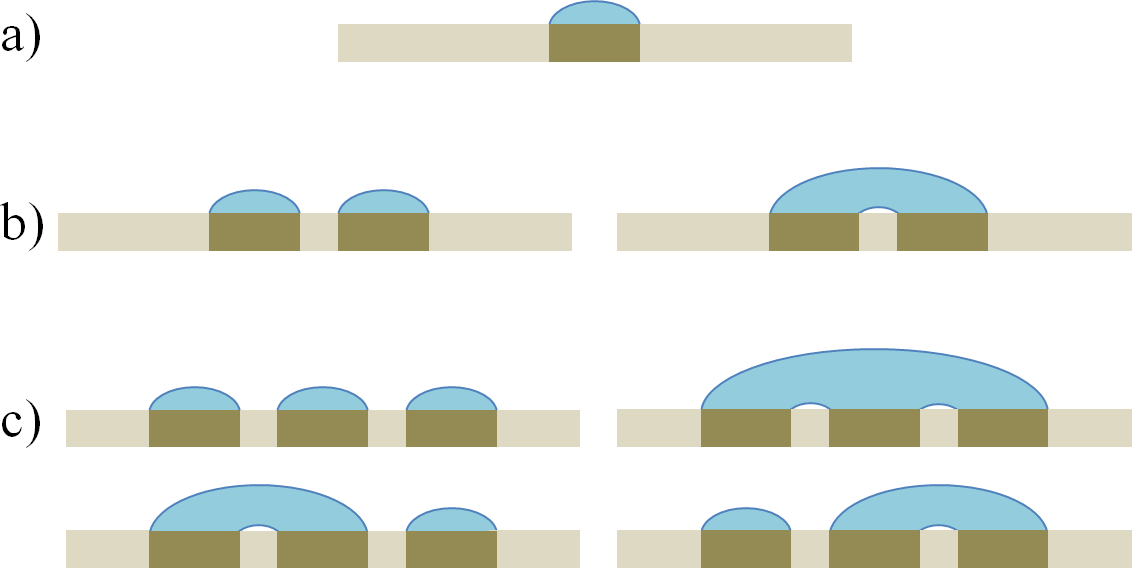}}
 \caption{Schematic illustration of possible isolated or bridging
 droplet configurations on a dry substrate ($\theta=\pi$) decorated with a) one b) two c) three completely wet stripes of width $L$}\label{fig1}
\end{figure*}

These scaling considerations are completely in accord with calculations based on a mesoscopic interfacial Hamiltonian model
 \bb
 H[h]=\int\dd x\left\{\frac{\gamma}{2}\left(\frac{d h}{d x}\right)^2+W(h)\right\}\,,\label{Hh}
 \ee
where $h(x)$ is the height of the drop above the stripe which extends from $x=-L/2$ to $x=L/2$ , $\gamma$ is the liquid-gas surface tension and
$W(h)$ is the binding potential. Translational invariance is assumed for the drop height along the stripe and the binding potential at bulk
coexistence is assumed to take the same form, $W(h)=A/h^2$ as for a uniform completely wet substrate where $A$ is the (positive) Hamaker constant.
This assumption is justified since the drop height is much smaller than the length $L$. Since the region outside the stripe is partially wet we may
impose that the drop height is fixed to a small, microscopic value at the edges $x=\pm L/2$. In three dimensions interfacial fluctuation effects at
complete wetting are near negligible and a mean-field treatment of interfacial model suffices to determine the drop shape. Minimization of (\ref{Hh})
leads to the Euler-Lagrange equation \cite{schick}
 \bb
  \gamma\frac{d^2 h}{d x^2}=W'(h)\,. \label{EL}
  \ee
which is easily solved for the drop shape. For wide stripes this leads to the scaling solution
   \bb
  h(x)=h_m\sqrt{1-4x^2/L^2}\,, \label{scale}
   \ee
where the mid-point height is determined as \cite{mpp}
  \bb
  h_m^2\approx L\sqrt{\frac{A}{2\gamma}}\,, \label{hm}
  \ee
in accord with the scaling prediction above. These mean-field results are not altered by the inclusion of interfacial fluctuations associated with
the thermal wandering of the interface. For example we can anticipate these lead a near negligible interfacial roughness at the mid-point which
scales as $\sqrt{\ln L}$ which is much smaller than the equilibrium height $h_m$. Evaluating $H[h]$ at the equilibrium profile shape determines that
the surface free-energy per unit length of this single drop behaves as
 \bb
  F_{\rm drop}(L)=\gamma L+\sqrt{2 A\gamma}\ln{L}+\cdots
  \label{Fsing}
 \ee
which indeed shows the logarithmic correction to the surface free-energy. The fact that this contribution diverges with $L$ means it is not possible
to define edge or line tension contributions to the surface free energy of the drop. This is very different to the case with purely short-ranged
forces for which the analogous finite-size is of order $1/L$. Returning to the case of long-ranged forces, taking a derivative w.r.t $L$ determines
that the force of solvation induced by the drop on the edges of the stripe is $f_{\rm sol}=\gamma+\mathcal{O}(1/L)$. The finite-size correction to
the surface tension here can be interpreted as a critical Casimir force induced by the complete wetting of the stripe which is characterized by a
large correlation length associated with interfacial fluctuations which is limited only by the value of the width $L$.

Next consider two completely wet stripes, each of width $L$, separated by a distance $D$. In this case two interfacial configurations are possible
(see Fig.~1b). If the distance $D$ is large the stripes are independent and isolated drops form on them. The excess surface free-energy associated
with this configuration is well approximated by $2F_{\rm drop}(L)+\gamma_{wg}D$. The last contribution here arises from the surface tension
associated with the wall-gas configuration above the region between the stripes. However if the distance $D$ is sufficiently small we may imagine
that a single drop bridges between the stripes with associated surface free-energy $F_{\rm drop}(2L)+(\gamma_{wl}+\gamma)D$. This approximation is
justified if $D\ll L$ which is indeed the case near the bridging transition. Matching these free-energies determines that these configurations
coexist when the width between the stripes is equal to
 \begin{equation}
 D_B=\frac{1}{1-\cos\theta}\sqrt{\frac{2A}{\gamma}}\ln L+\cdots
 \end{equation}
where recall $\theta$ is the contact angle above the non-wet parts away from the striped regions and the higher order terms remain finite as $L$
increases. Beyond the present mean-field analysis we anticipate that sharp, first-order bridging transitions are rounded due to fluctuation effects.
However the scale of the rounding is small, of order $\exp(-\gamma Dh_m/k_B T)$ which, even for microscopic systems, is negligible away from the bulk
critical temperature \cite{privman}.

Finally we turn attention to the case of three identical stripes separated by regions of width $D_1$ and $D_2$. To begin we consider the symmetric
situation $D_1=D_2=D$ which will be sufficient to determine the structure of the phase diagram. Depending on the width $D$ there are conceivably four
possible droplet configurations (Fig.~1c). If $D$ is large isolated drops cover the three wet stripes. From the considerations above it follows that
for $D<D_B$ two of the drops coalesce to form a bridge over one of the dry regions. Clearly there are two equivalent ways of doing this which break
the left/right symmetry of the substrate decoration (and underlying potential). However this transition may or may not be preceded by the coalescence
of all three drops into a single drop which bridges both dry regions, occurring at a distance $D_B^*$.

To determine $D_B^*$ we again simply compare the surface free-energy of a single bridging configuration with the value $3F_{\rm
drop}(L)+2\gamma_{wg}D$ corresponding to three isolated drops. Since we can anticipate that the value of $D_B^*\ll L$ we can approximate the
free-energy of the single bridging drop as $F_{\rm drop}(3L)+(2\gamma_{wl}+2\gamma)D$. This implies that the value of $D_B^*$ at which there is
coexistence of the three isolated drops and the single triple drop is given by
 \bb
  D_B^*=D_B+\frac{1}{1-\cos\theta}\sqrt{\frac{2A}{\gamma}}\ln\frac{2}{\sqrt{3}}\,.
 \ee
This result can also be obtained by solving the Euler-Lagrange equation for the full interfacial model and matching solutions at the boundaries
between the striped regions.

At this point we make a number of remarks:

i) As expected the values $D_B$ and $D_B^*$ diverge as the contact  angle $\theta$ vanishes, so that the entire wall is completely wet. Their values
are smallest for $\theta=\pi$, that is a dry wall, where there is the maximum penalty to the free-energy of bridging between the stripes. Although
they diverge with $L$ the scale of $D_B$ and $D_B^*$ are set by the magnitude of the prefactor $\sqrt{A/\gamma}$ which, away from the near vicinity
of the bulk critical temperature, is molecularly small. Thus in general bridging only occurs when the stripes are microscopically close to each
other.

ii) As $D_B^*>D_B$ the coalescence of the three isolated drops into a single drop as $D$ is reduced must precede the possible symmetry breaking
transition which, if present, must occur in a metastable region of the phase diagram. This is in keeping with the general expectation that the
equilibrium density profile of an inhomogeneous fluid must have the same symmetry as the confining external potential. It is intriguing however that
the difference $D_B^*-D_B$ is small and independent of $L$ implying that the symmetry breaking transition is only just metastable. As $L$ increases
the locations of the equilibrium (three drops to one) and metastable but symmetry breaking (three drops to a double drop) transition get relatively
closer to each other. Indeed in the next section we shall show that even for microscopic systems the metastable transition still lies very close to
the true equilibrium transition.

iii) Having determined that  $D_B^*>D_B$ it follows that when the distances $D_1$ and $D_2$ between the stripes are unequal the phase diagram should
have five first-order phase boundary lines and two triple points occurring for $D_1\neq D_2$ which lie close to each other. At each triple point
configurations corresponding to three isolated drops, a double bridging drop with an isolated drop and a single triple bridging drop all coexist. The
fact that $D_B^*>D_B$ rules out the alternative possibility the triple points occur along the diagonal $D_1=D_2$. The phase diagram will be
determined explicitly in the next section using a more microscopic model DFT.

\section{Microscopic Density Functional Theory}

\subsection{Microscopic Model}

Within DFT \cite{evans79}, the equilibrium state of an inhomogeneous fluid within the  grand-canonical ensemble is characterized by a one-body
density profile $\rhor$. In general, the equilibrium distribution is obtained by minimizing the grand potential functional
 \bb
 \Omega[\rho]={\cal F}[\rho]+\int\dd\rr\rhor[V(\rr)-\mu]\,,\label{om}
 \ee
where  $V(\rr)$ is the external potential. Here the intrinsic free energy functional ${\cal F}[\rho]$ can be separated into an exact ideal gas
contribution and an excess part:
  \bb
  {\cal F}[\rho]=\beta^{-1}\int\dr\rho(\rr)\left[\ln(\rhor\Lambda^3)-1\right]+{\cal F}_{\rm ex}[\rho]\,,
  \ee
where $\Lambda$ is the thermal de Broglie wavelength and $\beta=1/k_BT$. As is common in  modern DFT approaches, the excess part is modelled as a sum
of hard-sphere and attractive contributions where the latter is treated in a standard mean-field fashion \cite{archer_mf}:
  \bb
  {\cal F}_{\rm ex}[\rho]={\cal F}_{\rm hs}[\rho]+\frac{1}{2}\int\int\dd\rr\dd\rr'\rhor\rho(\rr')u_{\rm a}(|\rr-\rr'|)\,, \label{f}
  \ee
where  $u_{\rm a}(r)$ is the attractive part of the fluid-fluid interaction potential.

In our model the fluid atoms are assumed to interact with one another via the truncated (i.e., short-ranged) and non-shifted Lennard-Jones-like
potential
 \bb
 u_{\rm a}(r)=\left\{\begin{array}{cc}
 0\,;&r<\sigma\,,\\
-4\varepsilon\left(\frac{\sigma}{r}\right)^6\,;& \sigma<r<r_c\,,\\
0\,;&r>r_c\,.
\end{array}\right.\label{ua}
 \ee
which is cut-off at $r_c=2.5\,\sigma$, where $\sigma$ is the hard-sphere diameter.

The hard-sphere part of the excess free energy is approximated using the (original) fundamental measure functional (FMT) \cite{ros},
 \bb
{\cal F}_{\rm hs}[\rho]=\frac{1}{\beta}\int\dd\rr\,\Phi(\{n_\alpha\})\,,\label{fmt}
 \ee
which accurately takes into account the short-range correlations between the fluid particles. Here, $\{n_\alpha\}$ are six weighted densities
corresponding to fundamental measures of a sphere \cite{ros}. Although other prescriptions of $\Phi$ within modified versions of FMT are available
\cite{mulero}, the original Rosenfeld FMT functional is perfectly adequate to describe the packing effects in the system under study.

.

The external field $V(\rr)$ is exerted by a planar wall occupying the volume $z<0$. We specialize to the case $\theta=\pi$ for which recall the
predicted values of $D_B$ and $D_B^*$ are smallest. This is very easily achieved by assuming that the wall is purely repulsive, except for three deep
parallel stripes, each of width $L$. The total potential of the wall can thus be written as
 \bb
 V(x,z)=\left\{\begin{array}{cc}
 \infty\,;&z<0\,,\\
V_L(x,z)+V_L(x-D_1,z)&\\
+V_L(x-D_1-L-D_2,z)\,;&z>0\,,
\end{array}\right.\label{vext}
 \ee
where $V_L(x,z)$ is the potential due to a single stripe located at $0<x<L$ and $D_1$ and $D_2$ are the distances between the stripes. Here we also
assume that the stripes are macroscopically long, so that the system possesses translation invariance along the $y$-axis.

We suppose that the stripes atoms are distributed uniformly within their domains with a particle-density $\rho_w$ and interact with the fluid
particles via the Lennard-Jones $12$-$6$ potential
 \bb
 \phi(r)=4\varepsilon_w\left[\left(\frac{\sigma}{r}\right)^{12}-\left(\frac{\sigma}{r}\right)^{6}\right]\,.
 \ee
The net potential $V_L(x,z)$ is obtained by summing up the wall-fluid pair potential over the wall volume of the stripe. Therefore

\begin{eqnarray}
 V_L(x,z)=&&\rho_w\int_{x-L}^L\dd x'\int_{-\infty}^\infty\dd y'\int_z^\infty \dd z'\nonumber\times\\
            &&\times\phi\left(\sqrt{x'^2+y'^2+z'^2}\right)\,.
 \end{eqnarray}
 which can be expressed in the following scaling form
 \bb
  V_L(x,z)=\pi\varepsilon_w\rho_w\sigma^3\left[\frac{\sigma^{9}}{z^9}G_9\left(\frac{x}{z},\frac{L}{z}\right)
  -\frac{\sigma^{3}}{z^3}G_3\left(\frac{x}{z},\frac{L}{z}\right)\right]
 \ee
  with
 \bb
 G_9\left(\xi,\eta\right)=F_9(\xi-\eta)-F_9(\xi)
 \ee
 and
  \bb
 G_3\left(\xi,\eta\right)=F_3(\xi-\eta)-F_3(\xi)\,.
 \ee
   where
   \begin{widetext}
  \bb
  F_9(\xi)=\frac{2}{45}\left(1+\frac{1}{\xi^9}\right)-\frac{1}{2880}\frac {128\,{\xi}^{16}+448\,{\xi}^{14}+560\,{\xi}^{12}+280\,{
\xi}^{10}+35\,{\xi}^{8}+280\,{\xi}^{6}+560\,{\xi}^{4}+{\xi}^{2}+128}{{\xi}^{9} \left(1+{\xi}^{2}\right)^{7/2}}
  \ee
   \end{widetext}
   and
  \bb
  F_3(\xi)=\frac{1}{3}\left[1+\frac{1}{\xi^3}-{\frac {2\,{\xi}^{4}+{\xi}^{2}+2}{2{\xi}^{3} \sqrt {1+{\xi}^{2}}}}\right]\,.
  \ee

Minimization of (\ref{om}) leads to the Euler-Lagrange equation
 \bb
 V(\rr)+\frac{\delta{\cal F}_{\rm hs}[\rho]}{\delta\rho(\rr)}+\int\dd\rr'\rho(\rr')u_{\rm a}(|\rr-\rr'|)=\mu\,,\label{el}
 \ee
which we solve iteratively on a two dimensional grid $(0,X)\times(0,Z)$ where we set $X=200\,\sigma$ and $Z=30\,\sigma$. The first stripe is situated
within the interval $(x_i,x_i+L)$ where $x_i$ satisfies $2x_i+3L+D_1+D_2=X$. We impose the boundary conditions: $\rho(x,z_m)=\rho_b$ and
$\rho(0,z)=\rho(X,z)=\rho_\pi(z)$, where $\rho_b$ is the reservoir gas density and $\rho_\pi(z)$ is a 1D density profile of the fluid near a hard
wall.

\begin{figure}
\centerline{\includegraphics[width=9cm]{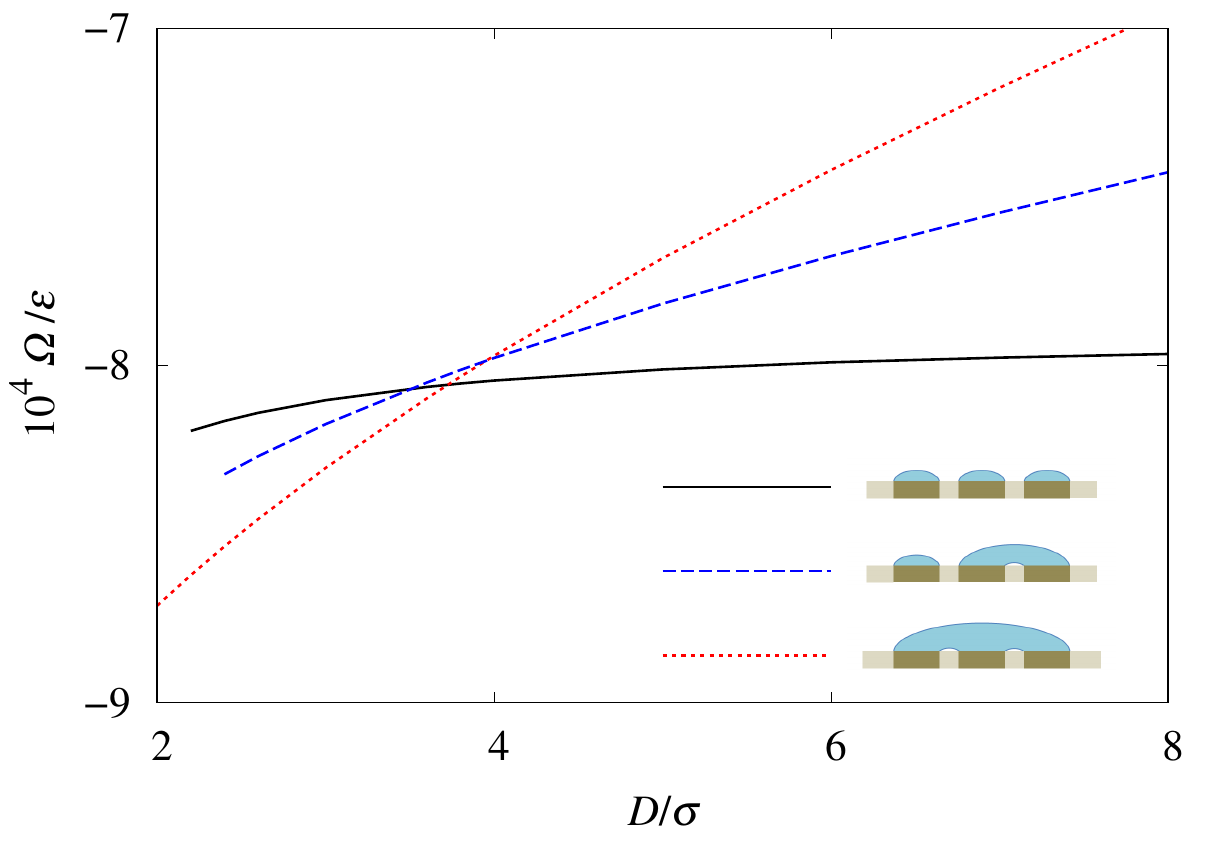}}
 \caption{DFT results showing the dependence of the grand potential on the stripe separation
$D$ for the symmetric situation when $D=D_1=D_2$. In keeping with the predictions of the interfacial model as the distance $D$ is reduced a single
bridging transition occurs from three isolated drops to a single triple droplet. Also shown is the grand potential of a metastable symmetry broken
configuration with one double droplet which lies close to the global minimum in the vicinity of the equilibrium bridging transition.}\label{om1}
\end{figure}

\begin{figure}
\centerline{\includegraphics[width=9cm]{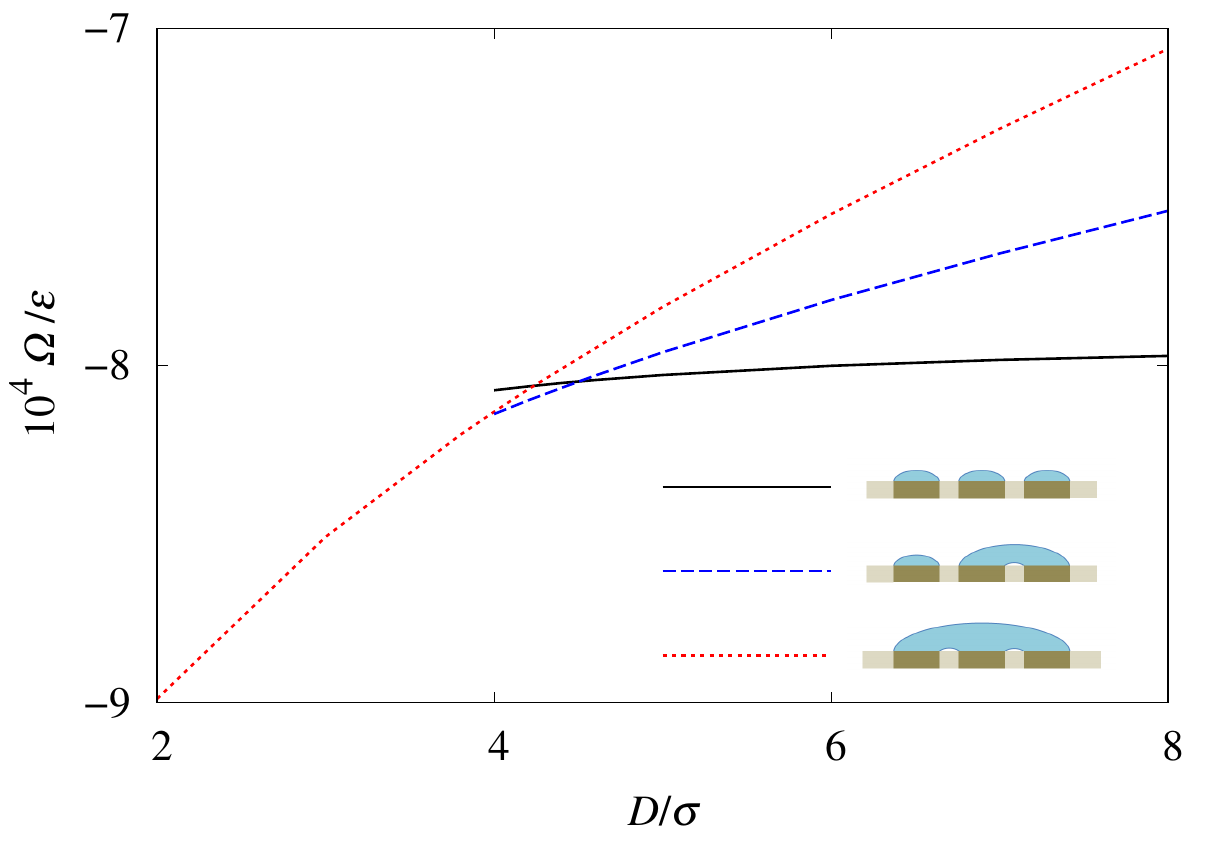}}
 \caption{Dependence of the grand potential on the separation $D=D_1=D_2-\sigma$ when the widths between the stripes are slightly different. In this case
  two bridging transition, first from three isolated drops to a double bridging drop and then to a triple bridging droplet occur as  $D$ is reduced. }\label{om2}
\end{figure}

\begin{figure*}
\centerline{\includegraphics[width=15.5cm]{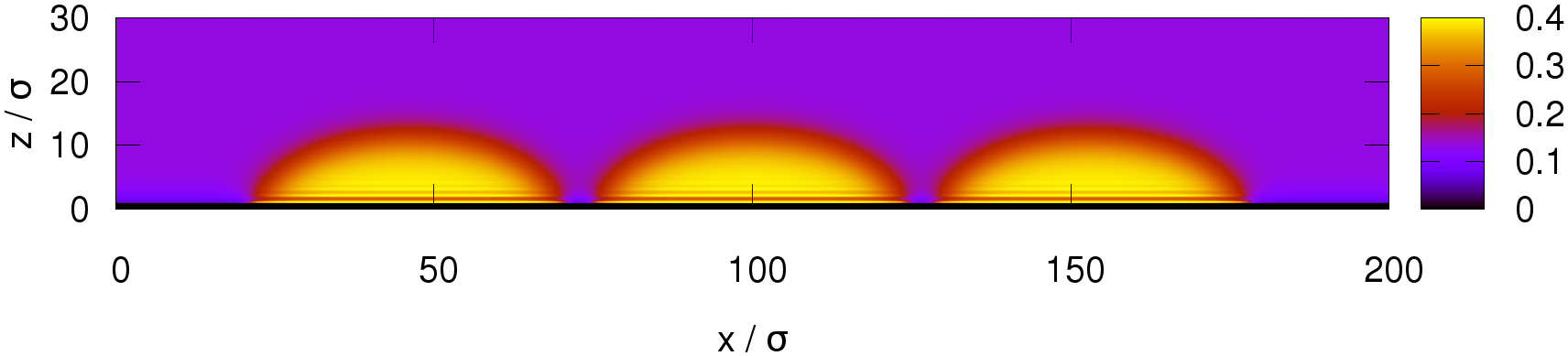}}
 \centerline{\includegraphics[width=15.5cm]{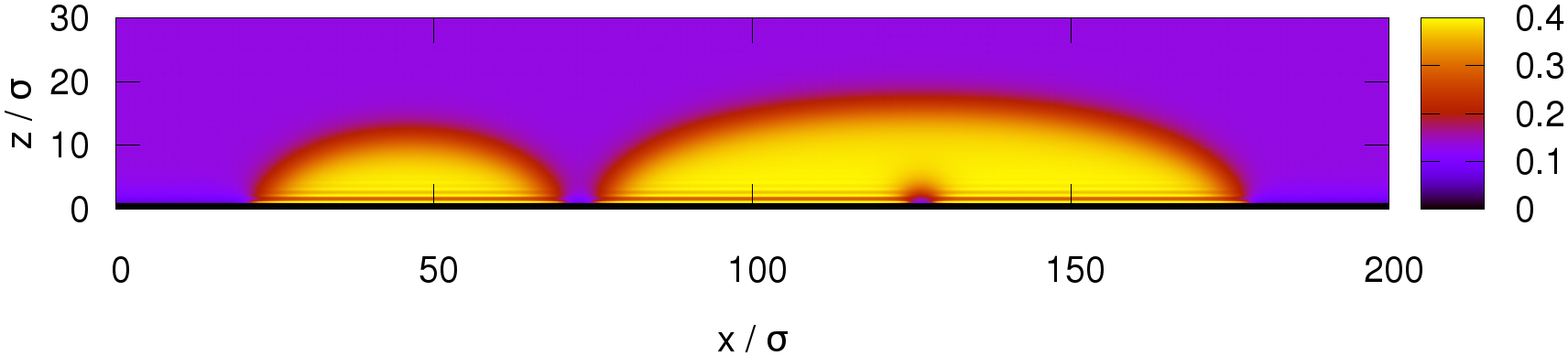}}
\centerline{\includegraphics[width=15.5cm]{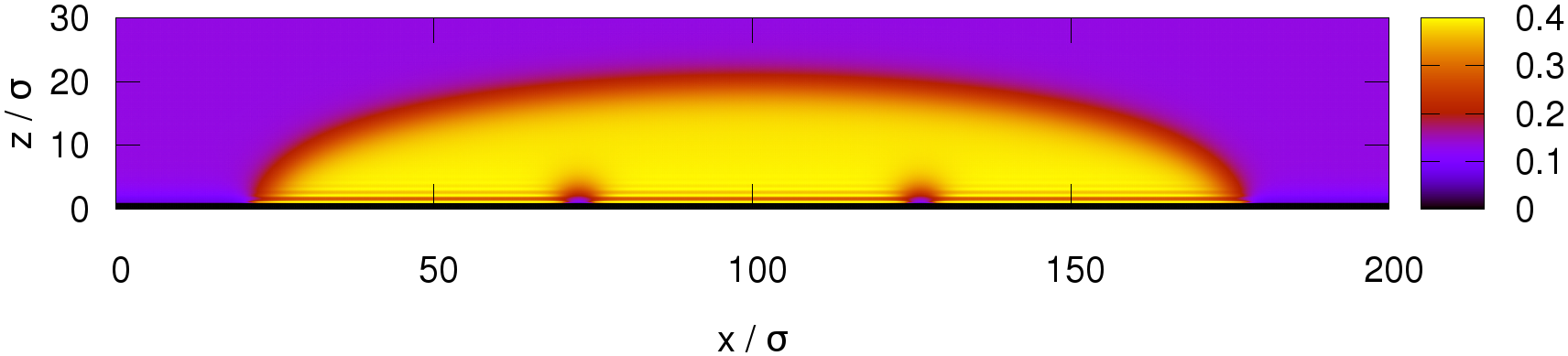}}
 \caption{DFT results for different equilibrium density profiles illustrating the microscopic structure underlying possible bridging morphologies of adsorbed liquid droplets.
Here the width of the wetting stripes is $L=50\,\sigma$ and the separations between them is (from top to bottom): i) $D_1=4\,\sigma$ and
$D_2=3.6\,\sigma$; ii) $D_1=4\,\sigma$ and $D_2=3.4\,\sigma$, and iii) $D_1=3.8\,\sigma$ and $D_2=3.6\,\sigma$. }\label{profs}
\end{figure*}

\subsection{DFT Results}
Within the present model the bulk critical temperature corresponds to $k_BT_c/\varepsilon=1.414$. In our calculations we choose a relatively large
value for the stripe potential, $\varepsilon_w\rho_w=\varepsilon\sigma^{-3}$, which at temperature $T=0.95\,~T_c$ ensures that the stripes are
completely wet \cite{wedge_mal}.

We next vary the distances between the stripes $D_1$ and $D_2$ and solve the Euler-Lagrange equation (\ref{el}) at bulk saturation chemical potential
 $\beta\mu_{\rm sat} =-2.94$ corresponding to bulk coexistence with associated densities $\rho_l=0.39\,\sigma^{-3}$ and
$\rho_g=0.13\,\sigma^{-3}$ of the liquid and gas bulk phases, respectively. Near a first-order bridging phase transition the density profile $\rhor$
obtained from Picard's iteration solution of the Euler-Lagrange depends on the initial configuration $\rho_i(\rr)$ of the fluid. If multiple
solutions are present that with the lowest grand potential corresponds to the equilibrium configuration. In order to converge to the isolated, double
and triple droplet configuration we chose: i) $\rho_i(\rr)=\rho_b\exp[-\beta V(\rr)]$ where the system is filled by a gas, ii) the same plus a slab
of liquid spanning two neighbouring stripes, iii) the same plus a slab of liquid spanning all three stripes. The height of the slabs was taken to be
approximately a square root of their width.

We first discuss the ``symmetric'' model wall with equally separated stripes, i.e. $D_1=D_2$. Fig.~\ref{om1} shows the dependence of the equilibrium
grand potential as a function of $D=D_1=D_2$. As can be seen, in this case the system undergoes a first order phase transition at $D\approx4\,\sigma$
from three droplets into one triple bridging droplet with two gas like bubbles inside above the dry regions. Also shown is a curve corresponding to
the grand potential of a metastable symmetry broken regime with two droplets. While this is never a global minimum of the grand potential the
different branches of the free-energy very nearly cross. This confirms the prediction that $D_B*>D_B$ and also that the distance between their values
is very small. In contrast, if we consider only slightly ``asymmetric'' model with $D=D_1=D_2-\sigma$, a double bridging droplet configuration can be
stabilized within a narrow interval around $D=4\,\sigma$ (see  Fig.~\ref{om2}). The equilibrium density profiles illustrating all four possible
droplet configurations are shown in Fig.~\ref{profs}.

\begin{figure*}
\centerline{\includegraphics[width=15cm]{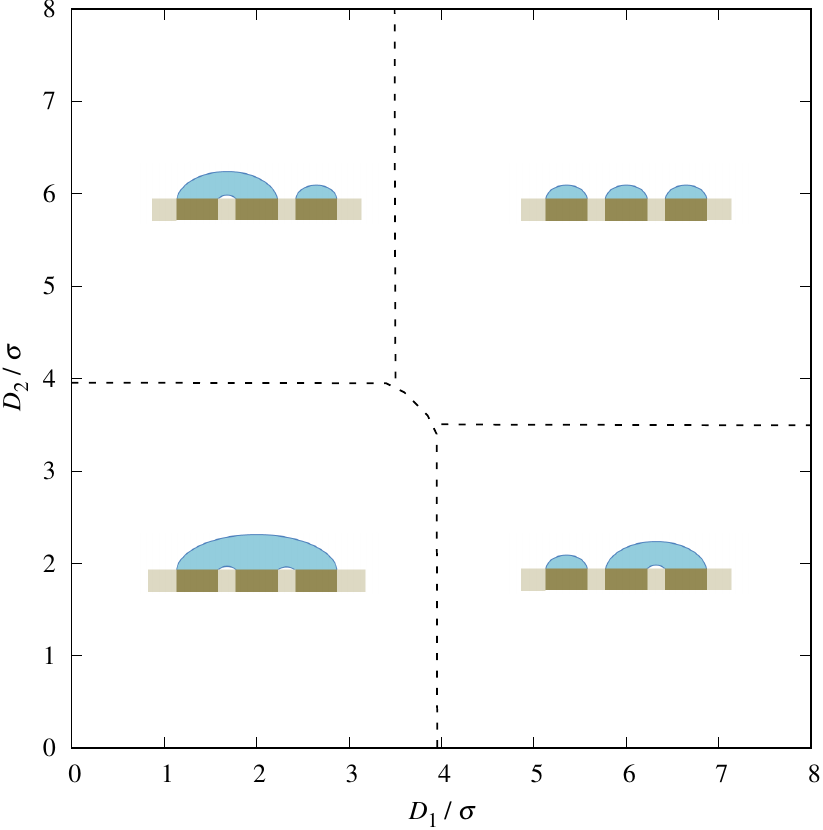}}
 \caption{Phase diagram in the $D_1$-$D_2$ plane showing the stable droplet configurations at bulk coexistence. Two triple points, occurring for
  $D_1\neq D_2$, connect five separate first-order phase boundaries. The triple points lie very close to each other reflecting the small difference
   between the equilibrium grand potential of the stable configuration and that of the metastable symmetry broken phase.
   }\label{pd}
\end{figure*}

Finally, in Fig.~\ref{pd} we show the phase diagram in the $D_1$--$D_2$ plane. To determine this we performed an extensive number of DFT calculations
by varying the values of $D_1$ and $D_2\ge D_1$ within the interval $D_1\in(1\,\sigma,10\,\sigma)$. One can see that the phase diagram separates into
four adsorption regimes with two triple points which occur away from the diagonal. Note, however, that the distance between the locations of the
triple points is less than one molecular diameter consistent with the prediction that $D_B^*-D_B$ is molecularly small. As is evident our microscopic
DFT results confirm the predictions of the mesoscopic interfacial Hamiltonian model for the phase diagram structure.  We are unable to verify the
logarithmic dependence on the width $L$ which would require much larger systems, nevertheless, the value of $D_B\approx 4\sigma$ is completely
consistent with the prediction (6) since  $A/\gamma\approx \sigma^2$. We stress again, however, that it is the difference between $D_B$ and $D_B^*$
rather than their absolute values which determines the topology of the phase diagram.

\section{Conclusion}

%(\emph{Perhaps into the conclusion part}): It should also be noted that our results imply that the system exhibits no symmetry breaking. It would
%only be in the case, if two configurations with two unlike droplets (one large and one small) coexisted. In this case the system would need to decide
%which configuration of the same free energy to choose. However, this coexistence is never realized since it would require the existence of the
%quadruple point. This result is important because it implies that for periodically decorated walls with an equidistant inter-stripe gap  one would
%not observe a sequence of first-order symmetry breaking transitions as anticipated in our recent study \cite{mpp} which would lead to extremely rich
%behaviour consisting of an infinite number of first-order phase transitions. The phase behaviour of a fluid adsorbed at such a wall will be very
%simple instead; there is only two stable configurations, one consisting of separated droplets adsorbed at each stripe and the other formed by a
%coalescence of all the droplets into a single one corresponding to a completely wet wall of macroscopic adsorption.

In this paper we have used a microscopic DFT model to study bridging transitions on chemically patterned surfaces. We have focused on systems with
long ranged dispersion forces where there is a strong finite-size correction to the droplet free energy. The results confirm the predictions of
simple interfacial Hamiltonian that  bridging only occurs when the distance between the stripes is microscopically small. When the distance between
the stripes $D_1$ and $D_2$ is varied the lines of phase coexistence intersect at two triple points. For the symmetric case $D_1=D_2\equiv D$ our
results also confirm that the droplet configuration always has the symmetry of the substrate corresponding to either three isolated drops when $D$ is
large or a single triple drop when $D$ is small. However, it is perhaps surprising that symmetry broken configurations occur in a metastable part of
the phase diagram which lies molecularly close to the equilibrium bridging phase transition as predicted consistently both by the Interfacial
Hamiltonian model and DFT.

It is natural to speculate that very similar results occur when the number of stripes is increased. In particular, when the distance between the
stripes is equal the only possible configurations correspond to isolated drops (when $D$ is large) or a single drop bridging across all the gaps
(when $D$ is microscopically small). As the number $N$ of stripes increases so does the height $h_m$ of the single bridging droplet; at bulk
coexistence we anticipate that $h_m\propto\sqrt{NL}$, since the size of the drop is approximately $NL$ in width (the dependence on $D$ may be dropped
as $D\ll L$). It follows that in the thermodynamic limit corresponding to a periodic array of completely wet and partially wet or dry stripes, that
at bulk coexistence, there are only two equilibrium configurations: If $D$ is large there is a bound phase corresponding to isolated drops which sit
above the wet regions. However, when $D$ is microscopically small, these coalesce causing the liquid-gas interface to unbind from the surface. The
simple interfacial Hamiltonian analysis of section II predicts that this transition occurs when $D$ takes a value
$D_B=\sqrt{\frac{2A}{\gamma}}\frac{\ln L}{(1-\cos\theta)}$ or equivalently when the fraction of the surface covered by the stripes is
$f=1-\mathcal{O}(\ln L/L)$. Since the liquid-gas interface unbinds from the substrate  at this point it follows that we may interpret this as a
first-order wetting transition induced by the chemical patterning of the substrate. The logarithmic dependence on $L$ here, induced by the strong
Casimir-like term in the drop free-energy, corresponds to a correction of the prediction based on Cassie's law which would require that the
transition occurs when $D=0$ or equivalently $f=1$.  Off coexistence this first-order wetting transition would also give rise to an associated
pre-wetting line corresponding to thin-thick transitions in the adsorption of fluid when $D<D_B$. We intend to study this prediction for the periodic
substrate using microscopic DFT in future work, as well as bridging transitions when drops are adsorbed on a surface decorated with circular patches
requiring a 3D analysis.

\begin{acknowledgments}
\noindent This work was funded in part by the EPSRC UK Grant no. EP/L020564/1, ``Multiscale Analysis of Complex Interfacial Phenomena". A.M.
acknowledges the support from the Czech Science Foundation, Project No. 17-25100S.
\end{acknowledgments}

\end{document}